# Performance Analysis and Comparison of Non-ideal Wireless PBFT and RAFT Consensus Networks in 6G Communications

Haoxiang Luo, *Graduate Student Member, IEEE*, Xiangyue Yang, Hongfang Yu, *Member, IEEE*, Gang Sun, *Senior Member, IEEE*, Bo Lei, Mohsen Guizani, *Life Fellow, IEEE*

*Abstract*—Due to advantages in security and privacy, blockchain is considered a key enabling technology to support 6G communications. Practical Byzantine Fault Tolerance (PBFT) and RAFT are seen as the most applicable consensus mechanisms (CMs) in blockchain-enabled wireless networks. However, previous studies on PBFT and RAFT rarely consider the channel performance of the physical layer, such as path loss and channel fading, resulting in research results that are far from real networks. Additionally, 6G communications will widely deploy high-frequency signals such as terahertz (THz) and millimeter wave (mmWave), while performances of PBFT and RAFT are still unknown when these signals are transmitted in wireless PBFT or RAFT networks. Therefore, it is urgent to study the performance of non-ideal wireless PBFT and RAFT networks with THz and mmWave signals, to better make PBFT and RAFT play a role in the 6G era.

In this paper, we study and compare the performance of THz and mmWave signals in non-ideal wireless PBFT and RAFT networks, considering Rayleigh Fading (RF) and close-in Free Space (FS) reference distance path loss. Performance is evaluated by five metrics: *consensus success rate*, *latency*, *throughput*, *reliability gain*, and *energy consumption*. Meanwhile, we find and derive that there is a maximum distance between two nodes that can make CMs inevitably successful, and it is named the *active distance* of CMs. The research results analyze the performance of non-ideal wireless PBFT and RAFT networks, and provide important references for the future transmission of THz and mmWave signals in PBFT and RAFT networks.

This work was supported by the Natural Science Foundation of Sichuan Province under Grant 2022NSFSC0913. An earlier version of this paper was presented in part at the IEEE International Conference on Metaverse Computing, Networking and Applications (MetaCom) 2023 [1]. (Corresponding author: Hongfang Yu, Gang Sun.)

Haoxiang Luo, and Gang Sun are with the Key Laboratory of Optical Fiber Sensing and Communications (Ministry of Education), University of Electronic Science and Technology of China, Chengdu 611731, China (e-mail: lhx991115@163.com; gangsun@uestc.edu.cn).

Xiangyue Yang is with the Pratt School of Engineering, Duke University, Durham, NC 27708, USA (e-mail: xy161@duke.edu).

Hongfang Yu is with the Key Laboratory of Optical Fiber Sensing and Communications (Ministry of Education), University of Electronic Science and Technology of China, Chengdu 611731, China, and also with Peng Cheng Laboratory, Shenzhen 518066, China (e-mail: yuhf@uestc.edu.cn).

Bo Lei is with the China Telecom Corporation Limited Research Institute, Beijing 100045, China (e-mail: leibo@chinatelecom.cn).

Mohsen Guizani is with the Machine Learning Department, Mohamed Bin Zayed University of Artificial Intelligence (MBZUAI), Abu Dhabi, United Arab Emirates (e-mail: mguizani@ieee.org).

## I. INTRODUCTION

Since the perfect combination of cryptography and consensus, blockchain is considered a revolutionary distributed system, which provides users with a decentralized architecture and strong tamper-proof capability. Blockchain is believed to have the potential to transform the way we share information and reshape society in the future. In the information and communication area, it is also expected to protect wireless network security in 6G communications [2-3]. Recently, it has been widely used in the Internet of Things (IoT) [4], Internet of Medical Things (IoMT) [5], Internet of Vehicles (IoV) [6], Internet of Drones (IoD) [7], and other network fields.

Therefore, it can be said that the emergence of blockchain paves the way for wireless networks in the future [8]. The CM in the blockchain is the basis that allows nodes in the network to establish trust without the involvement of any trusted third party. It plays an important role in reaching consistency in distributed systems. According to the design principle, CMs can be divided into two categories. The first is based on proof, including Proof of Work (Pow) [9], Proof of Stake (PoS) [10], Proof of Solution (PoSo) [11], and so on. And the second is based on voting, including PBFT [12], RAFT [13], Crash Fault Tolerance (CFT) [14], voting-based decentralized consensus (VDC) [15], and so on. Furthermore, there are also hybrid CMs such as delegated PoS-PBFT (DPOP) [16-17], and RAFT-enabled PBFT (RPBFT) or PBFT-enabled RAFT (PRAFT) [18].

### A. Research Motivation

PBFT and RAFT, as two typical voting-based CMs, are widely used in various scenarios due to their high fault tolerance and clear workflow. For PBFT, it has high throughput and low computational requirement, and provides 1/3 fault tolerance of the wireless network, that is, at most (*n*-1)/3 Byzantine nodes (*n* is the total number of nodes) are allowed in the system. For RAFT, it has a very simple workflow that avoids complex communication processes, and allows for (*n*-1)/2 failed nodes (but not Byzantine nodes) within the system. In a wireless network, where node failures are common due to the instability of the wireless environment, these two CMs can ensure the stability of the network. Therefore, studying the performance of PBFT and RAFT in wireless network scenarios can help us make further utilize them to achieve better negotiation results.

However, the current research on the performance of PBFT and RAFT wireless networks is quite ideal. For example, in [19], authors analyze the relationship between PBFT consensus success rate and delay in the ideal channel of IoV; and in [20], authors analyze the performance of wireless PBFT networks using IEEE 802.11 protocol. Additionally, [21] studies the RAFT performance in an Industrial IoT. Despite these researches, there are still several issues that need to be addressed to build a wireless network that supports blockchain. An important challenge is that wireless channels in the physical layer often suffer various channel fading and path losses. Channel fading can increase the bit error rate of signals, and path loss can hurt the received power of the receiving node, thus, they enhance the uncertainty of wireless connections and affect the overall performance of the blockchain. As a result, we need further to study and analyze the performance of non-ideal wireless PBFT and RAFT networks.

The path loss and channel fading models of wireless networks are not only related to the environment, but also to the signal frequency in the channel. In 6G communications, mmWave (26.5-100GHz) and THz (0.1-10THz) signals are regarded as important potential schemes because they provide more spectrum resources [1]. However, since their high frequencies, there are some problems such as large propagation attenuation and short transmission distance [22]. Their performances in various scenarios are worthy of further study. Therefore, the motivation of this paper is to investigate the performances of non-ideal wireless PBFT and RAFT networks considering THz and mmWave signals transmitted.

To sum up, three issues need to be addressed regarding the performance analysis of wireless PBFT and RAFT networks
- What is the performance of non-ideal wireless PBFT and RAFT networks carrying THz and mmWave signals in a 6G communications area? In particular, the core performance of CMs-- the *consensus success rate* deserves attention, because only the success of consensus can guarantee the consistency of the blockchain system.
- How does the *consensus success rate* impact the other key performances of wireless PBFT and RAFT networks in 6G wireless scenarios, such as the *consensus latency*, *throughput*, and *reliability gain*?
- *Energy consumption* is also one of the most important indicators for 6G communications and the blockchain system. It will be interesting to see how much power wireless PBFT and RAFT networks can generate in mmWave and THz environments.

*B. Our Contributions*

In this paper, the channel fading and path loss models we consider are RF and FS, respectively. Under their influence of them, we study the various processes of PBFT and RAFT consensus. Our work thus demonstrates the performance of various processes for non-ideal wireless PBFT and RAFT networks with THz and mmWave signals, and the factors that may affect their performance. To our best knowledge, this is the first work on performance analysis for THz and mmWave signals in PBFT and RAFT networks. The contribution of this paper can be summarized as follows
- First, we derive the performance of non-ideal wireless PBFT and RAFT networks with THz and mmWave signals, including *consensus success rate*, *latency*, and *throughput*. And the performance difference between these two CMs with two signals is compared by comprehensive simulations. These results provide a basic analytical range for non-ideal wireless PBFT and RAFT networks, and help to design a consensus beyond these two.
- Second, we take the logarithm of *consensus success rate* as *reliability gain*, and explore the relationship between it and the number of nodes. An amazing rule is that the relationship between the number of nodes and the *reliability gain* satisfies the Gaussian-like distribution. This rule will facilitate the deployment of PBFT and RAFT consensus in wireless networks.
- Third, we measure the *energy consumption* required for the consensus process. Meanwhile, we explore the relationship between *energy consumption* and other quality measures, including the number of nodes and *consensus success rate*. This practical concept provides an easy implementation of wireless networks in energy-constrained scenarios.
- Finally, we find a maximum distance between two nodes, named *active distance* of PBFT and RAFT. If the distance between any two nodes is less than this active value, these two CMs will inevitably succeed.

*C. Structure of This Paper*

The remaining contents of this paper are arranged as follows. Section II overviews the related work. Section III is the system model, which introduces the fundamentals of the PBFT and RAFT CM, as well as the RF and FS models. In Section IV, the *consensus success rate*, *latency*, *throughput*, and *energy consumption* of non-ideal wireless PBFT and RAFT networks with THz and mmWave signals are derived and analyzed in detail. Then, we compare the PBFT and RAFT consensus by simulating the above performances and *reliability gain*, and come to some interesting conclusions in Section V. Finally, we conclude this work in Section VI.

II. RELATED WORK

This section is divided into three parts, including research on PBFT consensus, research on RAFT consensus, and wireless networks.

*A. Research of PBFT Consensus*

Since the introduction of PBFT, there has been a lot of research on the performance analysis, optimization and application of PBFT.

For the performance analysis of PBFT, Li *et al*. [23] study the security of PBFT in the case of sharding, including two sharding models, which are non-cross sharding and cross sharding. In [24], authors analyze the performance of multilayer PBFT, including communication complexity and consensus success rate. In [25], authors model the PBFT consensus process using a random reward network to calculate the consensus time required for a 100-node network. Zheng *et al*. [26] use a continuous time Markov chain to simulate the time response of PBFT.

For the optimization and application of PBFT, Gang *et al*. [16], authors use delegated PoS (DPOS) to improve PBFT, to design a more efficient and promising CM, called DPOSP. In [18], authors use RAFT to improve PBFT with lower communication complexity and applied it to V2G networks. Xu *et al*. [26] optimize the PBFT process using a scoring and grouping mechanism, to meet the efficient communication need in the IoV. Additionally, PBFT is also commonly used in various networking scenarios, such as IoMT [5], energy-constrained IoT [27], and so on.

*B. Research of RAFT Consensus*

RAFT has also received a lot of attention due to its simple workflow. In this related work review of RAFT, we also divide it into performance analysis, optimization and application.

For the performance analysis of RAFT, Huang *et al*. [28] propose a simple analytical model to analyze the split probability of RAFT network, which is a function of network size, packet loss rate, and election time. In [29], authors implement RAFT on a software-defined network (SDN) plane and evaluate its data storage and leader selection performances.

For the optimization and application of RAFT, Luo *et al*. [18] use PBFT to improve RAFT, to make RAFT can allow the existence of Byzantine nodes. In [30], to reduce consensus latency, the authors offload some RAFT functionality onto a programmable P4-based switch. Wang *et al*. [31] optimize the leader election and consensus process of RAFT using the Kademlia protocol to improve leader election speed and throughput. In addition, RAFT is also used in a wide range of distributed systems, such as V2G networks [18], industrial IoT [21], high-speed networks [32], financial systems [33], and so on.

*C. Wireless Consensus Networks*

In recent work, there are few studies on CM in wireless communication scenarios. B. Cao *et al*. [34] analyze the performance of PoW, PoS and DAG blockchain in wireless networks, including consensus success rate, latency and throughput. Sun *et al*. [35-36] study a low-cost blockchain node deployment scheme in the wireless IoT. In [37], Herrera *et al*. use probabilistic broadcast scheduling to reduce the number of transmissions in wireless consensus networks. Xu *et al*. [38] design a fast fault-tolerant protocol for wireless networks, which have $n/2$ failed nodes.

Specifically, for PBFT and RAFT CM, they have also been studied in wireless networks. In [39], authors investigate the consensus latency and throughput required to operate a wireless PBFT network. Onireti *et al*. [40] study how to minimize the number of replicas to ensure PBFT consensus liveness in a wireless environment. In [41], authors use and improve PBFT in wireless sensor networks. And in [19], [42], authors investigate the performance of PBFT in wireless channels of IoV. Meanwhile, Xu *et al*. [43] study the number of malicious nodes in a wireless RAFT network. In [44], authors propose an improved two-stage RAFT consensus in a wireless environment. And in [45], authors research the PBFT and RAFT performance in the IoV environment.

Through the above literature review, we find that there are two problems in the existing research on wireless PBFT and RAFT networks: one is a lack of consideration of the channel characteristics in the physical layer; the other is not oriented to 6G communication networks research. And these two issues will be the focus of our paper.

III. SYSTEM MODEL

In this section, we introduce the PBFT CM, RAFT CM, the RF and FS models, respectively.

*A. PBFT Consensus*

Assuming that the wireless PBFT networks consist of *n* nodes, for a successful consensus, there should be no more than *b* Byzantine nodes, where *b* is related to *n* as follows.

$$b \leq \left\lfloor \frac{n-1}{3} \right\rfloor. \quad (1)$$

When *b* and *n* meet (1), the liveness of PBFT can be satisfied. However, when the total number of nodes is greater than $3b+1$, the performance of PBFT networks will not be improved, but the consensus efficiency will be reduced. Therefore, we assume that the number of nodes in the wireless PBFT networks satisfies $n=3b+1$.

Before the PBFT starts consensus, it selects the primary node through a process called view configuration, and the other nodes serve as replicas. According to the view change rule, each node in wireless PBFT networks may be selected as the primary node in turn. We denote the primary node as $v_m$, which can be denoted by

$$v_m = v(\mathrm{mod}\,|S|), \quad (2)$$

where, *S* represents the set of all nodes in the wireless PBFT networks, and *v* is the view number. Whenever the primary node fails, the view change rule is executed and a new primary node is selected.

After the primary node selection, the client sends a *request* to the primary node to enter the PBFT consensus process. In the case of functioning PBFT networks, a complete consensus process is divided into four stages: *pre-prepare*, *prepare*, *commit*, and *reply* (as shown in Fig. 1). Each node in the wireless PBFT networks should participate in the following consensus process.

- *Pre-prepare:* The primary node broadcasts a *pre-prepare* message to all replicas.
- *Prepare:* Each replica that receives the *pre-prepare* message broadcasts the *prepare* message to other replicas. If the replica receives $2b$ or more *prepare* messages corresponding to the *pre-prepare* message, this *prepare* message is considered valid.
- *Commit:* If the replica determines that the *prepare* message is true, it will broadcast the *commit* message to other replicas.

- *Reply:* Each replica returns a *reply* message to the client as a result of the reply message.

It is important to note that the result of the request is valid only if the client receives at least $b+1$ same *replies*.

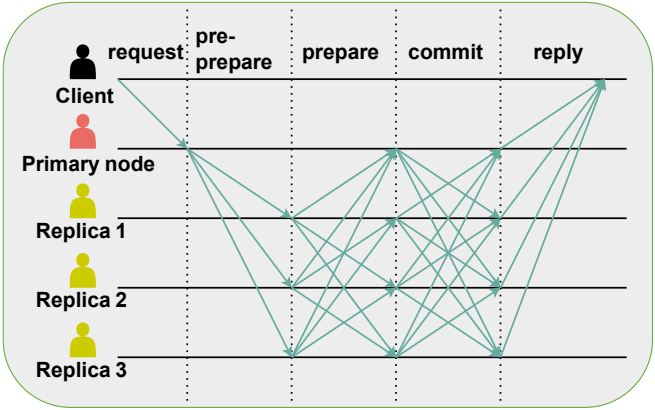

**Fig. 1.** PBFT consensus process.

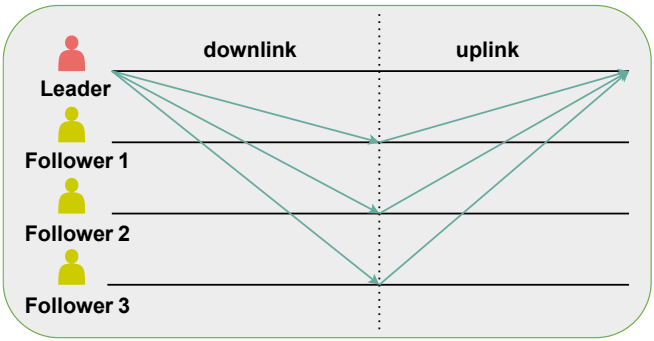

**Fig. 2.** RAFT consensus process.

*B. RAFT Consensus*

RAFT CM represents a network free of conflicts of interest. All nodes in the system are honest, and there are no Byzantine nodes. In such a network, the interaction of information is in the interest of each node [45].

Although there is no Byzantine node in the wireless RAFT network, there are still failed nodes. The wireless RAFT network is also assumed to have $n$ nodes. To ensure successful RAFT consensus, the number of failed nodes is at most $f$, and the following inequalities must be satisfied.

$$f \leq \left\lfloor \frac{n-1}{2} \right\rfloor. \quad (3)$$

Similar to PBFT, the optimal number of nodes for RAFT is (3) when the two sides are equal. Therefore, we assume that the number of nodes in the wireless RAFT networks satisfies $n=2f+1$.

Before starting the wireless RAFT network consensus, the leader node is elected by the node itself, and the other nodes are follower nodes. The subsequent consensus process is relatively simple, as shown in Fig. 2, which is divided into *downlink* communications and *uplink* communications. Each node in the wireless RAFT networks should participate in the following consensus process.

- *Downlink:* At this stage, the leader node broadcasts messages to all followers through *downlink* communications.
- *Uplink:* After followers receive a message from the leader, any node with the ability to judge will transmit its opinion to the leader via the *uplink*, confirming or denying it.

Only when the leader receives more than 50% confirmation messages, RAFT can be regarded as reached.

*C. RF and FS Models*

The characteristics of wireless communication channels determine the upper limit of the wireless communication system's performance. To make THz and mmWave signals better serve 6G communications, it is necessary to study the channel fading and path loss model.

First, we assume that nodes obey a two-dimensional Poisson distribution with density $\gamma$. Then, we randomly select a node as the sending node, taking it as the circle's center, and receiving nodes are distributed in the area with radius $R_a$. According to the two-dimensional Poisson distribution, the probability density function of the distance $r$ between the sending node and the receiving node can be expressed as

$$f(r) = \frac{d(r^2/R_a^2)}{dr} = \frac{2r}{R_a^2}. \quad (4)$$

When the channels in wireless PBFT and RAFT networks conform to RF, these channels are Rayleigh channels. Based on the characteristics of RF fading in wireless communications, the Signal-to-Interference-plus-Noise-Ratio (SINR) at the receiving node is

$$SINR = \frac{P_T h r^{-\alpha}}{P_N + P_I}, \quad (5)$$

where, $P_T$ is the node's transmit power; $h$ represents a non-negative random variable of power gain in RF, which follows a negative exponential distribution with exponent 1. $\alpha$ represents the path loss exponent; $P_N$ and $P_I$ are the noise and interference power. We find that the parameters in (5) are all constant except $\alpha$. $\alpha$ is a variable related to the path loss. Therefore, in the next step, we need to analyze the path loss model for THz and mmWave signals.

Second, we assume that the path loss model for THz and mmWave signals is FS. According to [46], the path loss on a specific distance can be expressed as the logarithmic distance

$$PL(r)_{av} = PL(r_0) + 10\alpha \log\left(\frac{r}{r_0}\right) + X_\sigma, \quad (6)$$

where $PL(r)_{av}$ is the average path loss at distance $r$; $PL(r_0)$ represents the path loss at reference distance $r_0$ according to the FS model; $X_\sigma$ is a zero-mean Gaussian random variable with standard deviation $\sigma$.

In this paper, we adopt path-loss exponent ($\alpha$=2.229) with 0.22THz signals in [46], and path-loss exponent ($\alpha$=1.7) with 28GHz signals in [47-48], respectively.

IV. PERFORMANCE ANALYSIS

In this section, we analyze the *consensus success rate*, *latency*, *throughput*, and *energy consumption* of wireless PBFT and RAFT networks respectively.

*A. PBFT Consensus*

  *1) Consensus Success Rate*

We set the SINR threshold at which nodes can recover signals as $z$, then according to the two-dimensional Poisson distribution [49-50], the average success probability of transmission is

$$P_s = \int_0^R P\{SINR > z\} f(r) dr$$
$$= \frac{2\pi\gamma}{n} \int_0^{\sqrt{n/(\pi\gamma)}} \exp\left\{\frac{-(P_N + P_I)r^\alpha z}{P_T}\right\} r dr. \quad (7)$$

After the PBFT completes the view configuration, the communication will be divided into four steps: *pre-prepare*, *prepare*, *commit*, and *reply*, thus, we analyze the success rate of these four steps in turn.

**Success rate of pre-prepare:** At this stage, after receiving a *request* from a client, the primary node broadcasts the *pre-pare* message to every replica. According to the fault tolerance of PBFT, this stage allows a maximum of $b$ communication failures. Therefore, the success rate of *pre-prepare* is

$$P_{pre-prepare} = \sum_{i=0}^{b} C_{n-1}^i (1-P_s)^i P_s^{(n-1-i)}. \quad (8)$$

**Success rate of prepare:** Given the success rate (8) at the *pre-prepare* stage, $n$-1-$i$ nodes receive the *pre-prepare* message from the primary node. Then, each replica broadcast a *prepare* message to other replicas. To ensure successful completion of this stage, a maximum of $b$-$i$ communication failures are allowed, because the *pre-prepare* stage has $i$ failures. Then, the success rate of *prepare* is

$$P_{prepare} = \sum_{j=0}^{b-i} C_{n-1-i}^j (1-P_s)^j P_s^{(n-1-i-j)}. \quad (9)$$

**Success rate of commit:** This stage is very similar to prepare. The only difference is that the primary node also needs to broadcast a *commit* message. After the first two phases, $i+j$ nodes have not received messages properly. Therefore, the success rate of *commit* is

$$P_{commit} = \sum_{k=0}^{b-i-j} C_{n-i-j}^k (1-P_s)^k P_s^{(n-i-j-k)}. \quad (10)$$

**Success rate of reply:** To ensure that the *reply* stage is valid, the client must receive $2b+1$ *reply* messages. In other words, this stage allows a maximum of $b$-$i$-$j$-$k$ communication failures, thus the success rate of the *reply* stage is

$$P_{reply} = \sum_{l=0}^{b-i-j-k} C_{n-i-j-k}^l (1-P_s)^l P_s^{(n-i-j-k-l)}. \quad (11)$$

In general, the consensus success rate of wireless PBFT networks is closely related to (8), (9), (10) and (11), and can be expressed as (12).

*2) Latency and Throughput*

Based on the above analysis of PBFT consensus, its communication is divided into four stages, thus, the latency of wireless PBFT networks is the sum of the four-stage communication latencies. In each stage, the relationship between the communication latency and $P_s$ can be known from [19], [21] and [51], namely

$$1 - P_s = f_Q\left(\frac{NTBC - NTBR_T + \frac{\log NTB}{2}}{(\log e)\sqrt{NTB}}\right), \quad (13)$$

where, $f_Q$ is the $Q$ function, and $T$ represents latency for a channel. $N$ represents the number of subcarriers, in our paper $N = 1$. $B$ is the bandwidth. $R_T$ and $C$ are the transmission rate and channel capacity, respectively.

For the *pre-prepare* stage, the primary node needs to send messages to $n$-1 replicas, thus, the communication latency $t_{pre-prepare}$ can be expressed as

$$t_{pre-prepare} = (n-1)T. \quad (14)$$

Similarly, in the *prepare* and *commit* stages, each node needs to broadcast messages to $n$-1 nodes, thus, $t_{prepare}$ and $t_{commit}$ are consistent with $t_{pre-prepare}$, which can be represented by $t_1$.

$$t_1 = t_{pre-prepare} = t_{prepare} = t_{commit} = (n-1)T. \quad (15)$$

In addition, in the *reply* stage, each node only needs to send a *reply* message to the primary node, thus, its latency $t_{reply}$ is equal to $T$, which can be represented by $t_2$.

$$t_2 = t_{reply} = T. \quad (16)$$

According to (12), we can calculate $t_1$ and $t_2$, respectively, namely (16) and (17).

$$1 - P_s = f_Q\left(\frac{\frac{t_1}{n-1}B(C - R_T) + \frac{1}{2}\log\left(\frac{t_1}{n-1}B\right)}{(\log e)\sqrt{\frac{t_1}{n-1}B}}\right), \quad (17)$$

$$1 - P_s = f_Q\left(\frac{t_2 B(C - R_T) + \frac{1}{2}\log(t_2 B)}{(\log e)\sqrt{t_2 B}}\right). \quad (18)$$

Furthermore, we can obtain that the total latency of the wireless PBFT network is

$$t_P = t_3$$
$$= 3t_1 + t_2. \quad (19)$$

Moreover, from the definition of throughput *TPS*, we know that throughput is the number of transactions generated per second. And the transaction generation depends on reaching consistency, thus, the throughput can be represented by the consensus time, that is, the inverse of the total latency.

$$TPS_P = \frac{1}{t_P}. \quad (20)$$

---

$$P_P = \sum_{i=0}^{f} \left( C_{n-1}^i (1-P_s)^i P_s^{(n-1-i)} \sum_{j=0}^{f-i} \left( C_{n-1-i}^j (1-P_s)^j P_s^{(n-1-i-j)} \sum_{k=0}^{f-i-j} \left( C_{n-i-j}^k (1-P_s)^k P_s^{(n-i-j-k)} \sum_{l=0}^{f-i-j-k} C_{n-i-j-k}^l (1-P_s)^l P_s^{(n-i-j-k-l)} \right) \right) \right), \quad (12)$$

$$P_R = \sum_{i=0}^{f} \left( C_{n-1}^i (1-P_s)^i P_s^{(n-1-i)} \sum_{j=0}^{f-i} C_{n-1-i}^j (1-P_s)^j P_s^{(n-1-i-j)} \right). \quad (28)$$

where $P_s$ can be obtained by (6).

TABLE I
PARAMETER VALUES

| Parameters | | Values |
|---|---|---|
| THz signal | $P_N+P_I$ | 0.2 W |
| | $P_T$ | 1 W |
| | B | 10 GHz |
| | C | 80 Gbps |
| | R | 40 Gbps |
| | α | 2.229 [46] |
| mmWave signal | $P_N+P_I$ | 0.2 W |
| | $P_T$ | 1 W |
| | B | 800 MHz |
| | C | 8 Gbps |
| | R | 4 Gbps |
| | α | 1.7 [47-48] |

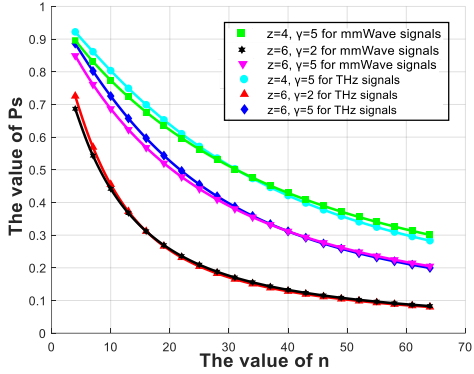

**Fig. 3.** The value of $P_s$.

*3) Energy Consumption*

The simplest way to calculate energy consumption is power times time ($P_T$* *latency*), thus, we only need to calculate the energy consumption of each stage in the wireless PBFT network, and then sum up to get the total energy consumption required by the PBFT to reach the consistency.

For the *pre-prepare* stage, there exist *n*-1 communications, so the energy consumption at this stage can be expressed as

$$E_{pre-prepare} = (n-1)t_1 P_T. \quad (21)$$

For the *prepare* stage, each replica broadcasts a *prepare* message to the other nodes, so there exist (*n*-1)² communications. And the energy consumption is

$$E_{prepare} = (n-1)^2 t_1 P_T. \quad (22)$$

For the *commit* stage, each node, including all replicas and the primary node, broadcasts a *commit* message to the other nodes, so there exists *n*(*n*-1) communications. Then, the energy consumption at this stage is

$$E_{commit} = n(n-1)t_1 P_T. \quad (23)$$

For the *reply* stage, there exist *n* communications, so the energy consumption can be expressed as

$$E_{reply} = nt_2 P_T. \quad (24)$$

To summarize, we sum the energy consumption of the four stages above, and the total energy consumption for the wireless PBFT network is

$$E_P = E_{pre-prepare} + E_{prepare} + E_{commit} + E_{reply} \\ = (2n^2 t_1 - 2nt_1 + nt_2)P_T. \quad (25)$$

*B. RAFT Consensus*

*1) Consensus Success Rate*

When the RAFT completes the leader node election, the *downlink* communications and *uplink* communications are conducted respectively. Therefore, the consensus success rate is calculated separately according to these two stages. Similar to PBFT, we use the same $P_S$ value.

**Success rate of downlink:** At this stage, the leader node transmits the *leader* message to every follower. According to the 50% fault tolerance of RAFT, the maximum allowable communication failure at this stage is *f*. Therefore, the success rate of the *downlink* is

$$P_{downlink} = \sum_{i=0}^{f} C_{n-1}^{i} (1-P_s)^i P_s^{(n-1-i)}. \quad (26)$$

**Success rate of uplink:** After completing the *downlink* stage, *n*-1-*i* nodes receive the *leader* message. Then, each follower replies the judgment to the *leader* via *uplink* communications. To ensure a successful RAFT consensus, the maximum allowable communication failure is *f*-*i*, because the *downlink* stage has *i* failures. Then, the success rate of the *uplink* is

$$P_{uplink} = \sum_{j=0}^{f-i} C_{n-1-i}^{j} (1-P_s)^j P_s^{(n-1-i-j)}. \quad (27)$$

In general, the consensus success rate of wireless RAFT networks is closely related to (26) and (27), and can be expressed as (28).

*2) Latency and Throughput*

Similar to PBFT, the consensus latency for wireless RAFT networks can also be calculated by (13). For the *downlink* communications, the leader sends the *lead* message to *n*-1 followers, thus, the latency at this stage can be expressed by $t_1$. And for the *uplink* communications, each follower only needs to send its judgment to the leader node, thus, the latency at this stage can be represented by $t_2$. Then, the total latency of the wireless RAFT network is

$$t_R = t_4 \\ = t_1 + t_2. \quad (29)$$

And the transaction throughput can be represented by the inverse of the total latency.

$$TPS_R = \frac{1}{t_R}. \quad (30)$$

*3) Energy Consumption*

The energy required to achieve consistency in a wireless RAFT network can be expressed as the sum of the energy consumption of *downlink* communications and *uplink* communications.

For the *downlink* stage, there exist *n*-1 communications, so the energy consumption at this stage can be expressed as

$$E_{downlink} = (n-1)t_1 P_T. \quad (31)$$

And for the *uplink* stage, there also exists *n*-1 communications, so the energy consumption at this stage can be expressed as

$$E_{uplink} = (n-1)t_2 P_T. \quad (32)$$

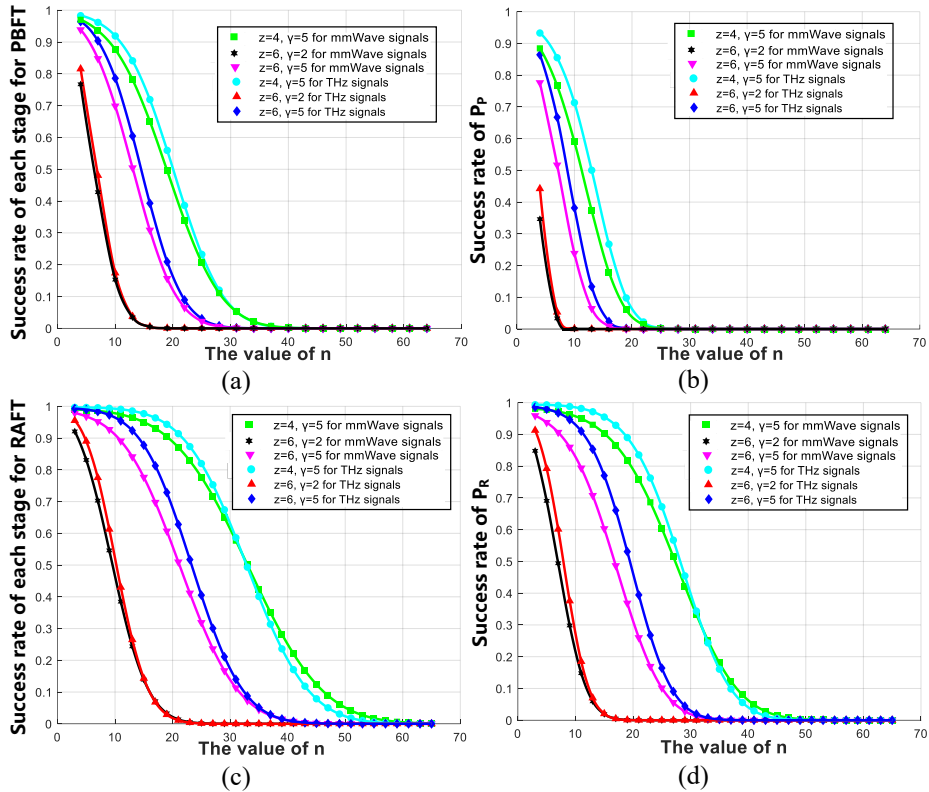

**Fig. 4.** (a) Average success rate of each stage for PBFT; (c) The value of $P_P$; (c) Average success rate of each stage for RAFT; (d) The value of $P_R$.

In summary, the total energy consumption for the wireless RAFT network is

$$E_R = E_{downlink} + E_{uplinnk} \\ = (n-1)(t_1 + t_2)P_T. \quad (33)$$

## V. PERFORMANCE SIMULATION

Before simulations, we need to set the parameter values related to the above performances in wireless PBFT and RAFT networks, which are shown in Table I. In addition, in order to study the effects of the SINR threshold $z$, and node density $\gamma$ on the three performances, we assume three sets of data to get a more comprehensive reference, which are $z$=6 dB, $\gamma$=2 nodes/m²; $z$=6 dB, $\gamma$=5 nodes/m²; $z$=4 dB, $\gamma$=5 nodes/m².

### A. Simulation of Success Rate

For the success rate, first, we simulate the transmission success rate of THz and mmWave signals in RF and FS models, namely (7). The simulation results are shown in Fig. 3, showing the relationship between $P_s$ and $n$. As the number of nodes increases, $P_s$ will decrease. The reason is that, according to the two-dimensional Poisson distribution, when there are too many nodes in wireless networks, the distance between some nodes becomes too large. Then, with the increase of distance, the influence of RF and FS is more obvious, thus, the $P_s$ value is smaller. Additionally, the values of $z$ and $\gamma$ also affect the $P_s$ values. A lower value of $z$ indicates that the receiving node has a stronger capability to recover the signal, which contributes to the transmission success rate. The lower value of $\gamma$ indicates that the distance between nodes increases, leading to a decrease in the transmission success rate. Moreover, under the same $z$ and $\gamma$ values, when the number of nodes is small, the performance of mmWave signals is worse than THz signals. And when the number of nodes increases, the performance of mmWave signals is better than THz signals. The reason for this is that THz is a higher frequency signal than mmWave, and is more likely to be negatively affected by distance.

Second, we simulate the average success rate of each stage in wireless PBFT networks. Due to the success rate of later stages depending on earlier stages, we represent the average of four stages as the success rate of each stage. Then, the results are shown in Fig. 4 (a). When the number of nodes is small, the fault tolerance of PBFT consensus can improve the transmission success rate compared with Fig. 3. However, when the $n$ value is large, the transmission success rate significantly decreases, indicating that the increase of distance between nodes has a negative impact on the communication ability of wireless PBFT networks. In addition, under the same $z$ and $\gamma$ values, the success rate of THz signals is higher than that of mmWave signals.

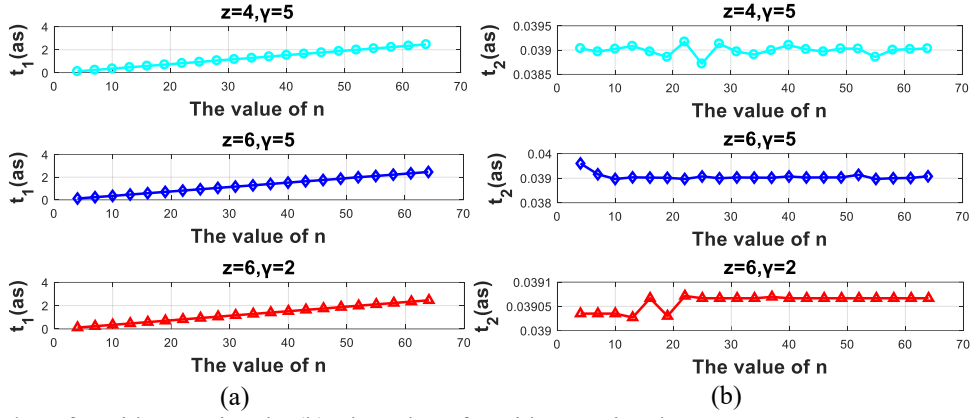

**Fig. 5.** (a) The value of $t_1$ with THz signals; (b) The value of $t_2$ with THz signals;

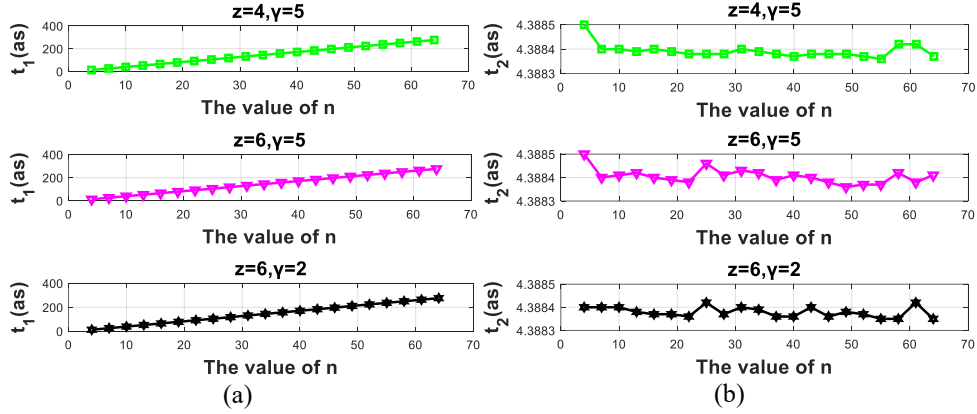

**Fig. 6.** (a) The value of $t_1$ with mmWave signals; (b) The value of $t_2$ with mmWave signals.

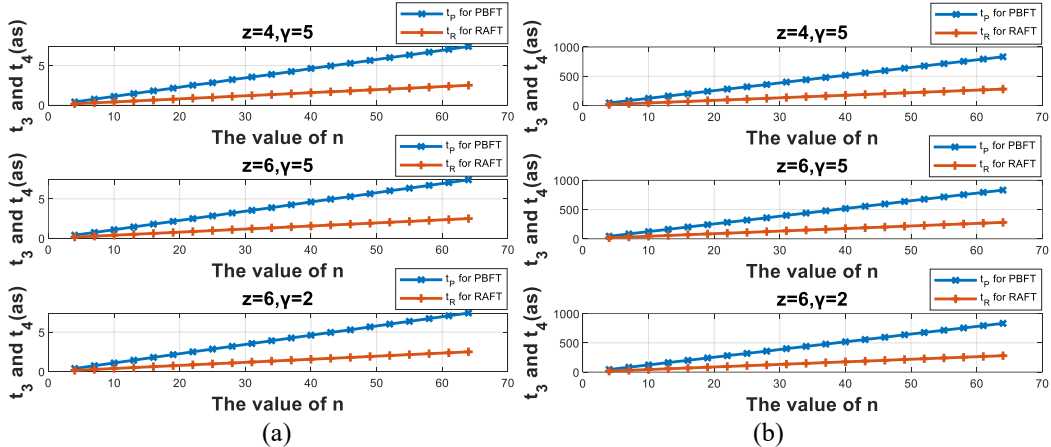

**Fig. 7.** (a) The value of $t_P$ and $t_R$ with THz signals; (b) The value of $t_P$ and $t_R$ with mmWave signals.

Third, we simulate the consensus success rate $P_P$ of wireless PBFT networks, as shown in Fig. 4 (b). This result is similar to Fig. 4 (a), except that the value of $P_P$ is decreased compared with Fig. 4 (a). This result is quite understandable that $P_P$ is the probability of simultaneous success in all four stages of PBFT consensus, while Fig. 3 (b) is the success rate of one stage. Additionally, the decrease of $P_P$ value with the increase of $n$ indicates that mmWave and THz signals are not suitable for communication in long-distance wireless PBFT networks. This result just shows that high-frequency signals such as mmWave and THz are affected by spatial distance easily. Most importantly, we can find that the consensus success rate of THz signals in wireless PBFT networks is higher than that of mmWave signals. This indicates that a bigger $\alpha$ value has a better consensus success rate in the two-dimensional Poisson distribution, when the value of $n$ is not large enough.

Finally, we simulate the average success rate of each stage and the consensus success rate of wireless RAFT networks, as shown in Fig. 4 (c) and (d), which are roughly similar to PBFT and will not be described in detail. However, there is a significant and attractive difference. The RAFT value is greater than PBFT for both the success rate of each stage and the consensus success rate. The reason for this is that RAFT has a higher fault tolerance rate than PBFT, allowing up to 50% of node failures compared to 33% for PBFT. In addition, for the consensus success rate, RAFT is the product of two communication stages, and PBFT is the product of four communication stages, thus, the consensus success rate of RAFT will be naturally greater than that of PBFT.

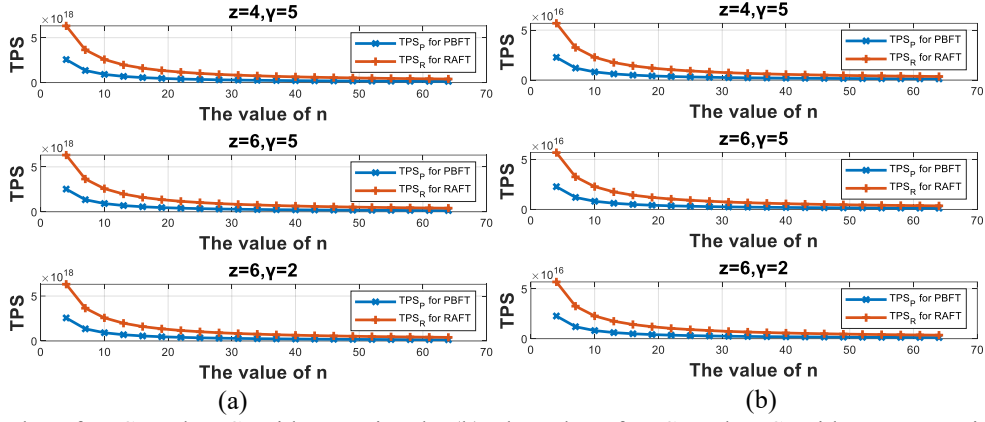

**Fig. 8.** (a) The value of $TPS_P$ and $TPS_R$ with THz signals; (b) The value of $TPS_P$ and $TPS_R$ with mmWave signals.

After the above simulation about success rate, we find that on the one hand, the SINR threshold $z$ of the receiving node can be reduced to improve the success rate; On the other hand, it can improve the node density to increase the success rate. Moreover, we also find that the distance between nodes is an important factor affecting the transmission success rate of wireless PBFT and RAFT networks, thus, we hope to explore a maximum distance to make CMs inevitably successful. As a result, further to ensure that the receiving node can recover the signal, its SINR threshold $z$ should be less than or equal to the SINR of signals, namely

$$z \le SINR = \frac{P_T h r^{-\alpha}}{P_N + P_I}. \tag{34}$$

Then, we can obtain the relationship between $r$ and other parameters in (35).

$$r \le \left(\frac{P_T h}{z(P_N + P_I)}\right)^{-\alpha}. \tag{35}$$

If the distance between any two nodes in wireless consensus networks satisfies (35), then the CM must be successful. And this distance is named the *active distance* of CM.

*B. Simulation of Latency*

First, we simulate $t_1$, $t_2$, and $t_{total}$ with THz and mmWave signals, separately.

For **THz signals**, Fig. 5 (a) shows the relationship between $t_1$ and the number of nodes $n$ with different values of $z$ and $\gamma$. $t_1$ shows a linear growth trend with the increase of $n$, and the values of $z$ and $\gamma$ have little influence on the change of $t_1$. This indicates that broadcasting messages to multiple nodes at the same time can prolong the latency. Fig. 5 (b) shows the relationship between $t_2$ and $n$ with different values of $z$ and $\gamma$. $t_2$ does not show a regular change with the increase of $n$, but shows a fluctuation characteristic. Meanwhile, with different values of $z$ and $\gamma$, the fluctuation law of $t_2$ is not completely consistent. And the fluctuation range always stays at 0.038-0.04as (1as=10$^{-18}$s). This result indicates that $t_2$ is one or two orders of magnitude smaller than $t_1$, indicating that $t_1$ plays a decisive role in the total latency. As a result, for PBFT, the latency of the first three stages (*pre-prepare*, *prepare*, *commit*) accounts for most of the total latency, while the latency of the last stage (*reply*) is so short that it can be ignored. And for RAFT, the latency of the *downlink* communications accounts for most of the total latency, while the latency of the *uplink* communications is so short that it can be ignored. In Fig. 7 (a), $t_P$ and $t_R$ show a changing trend similar to $t_1$. They grow linearly with the increase of $n$. Their values are in the order of as, which show the high-speed character of the THz signal. Additionally, due to the more complex communication process of PBFT, the consensus latency is nearly three times that of RAFT. Furthermore, when THz signals are transmitted fast in wireless PBFT and RAFT networks the simulation results also show that the number of nodes plays an important role in the latency.

For **mmWave signals**, Fig. 6 (a), (b), and Fig.7 (b) show the characteristics of $t_1$, $t_2$, and $t_P$, $t_R$, respectively. $t_1$, $t_P$, and $t_R$, have similar properties to THz signals, but two orders of magnitude more than THz signals. This indicates that THz can provide more bandwidth than mmWave, thus, it has higher communication rates. Additionally, $t_2$ also shows an irregular fluctuation characteristic, and the fluctuation range is from 4.3883-4.3885as.

Second, we simulate the throughput of wireless PBFT and RAFT networks under mmWave and THz signals, respectively.

For **THz signals**, Fig. 8 (a) shows the relationship between TPS and the number of nodes $n$ with different values of $z$ and $\gamma$. We can find that with the increase of $n$, TPS shows a decreasing trend, which can be approximated as an empirical decreasing curve. Despite TPS having a downward trend, it is consistently in the order of $10^{17}$-$10^{18}$ in our simulation, which indicates that wireless PBFT networks with THz signals have a very high throughput. Additionally, the throughput of RAFT is nearly three times that of PBFT, because RAFT has a lower consensus latency.

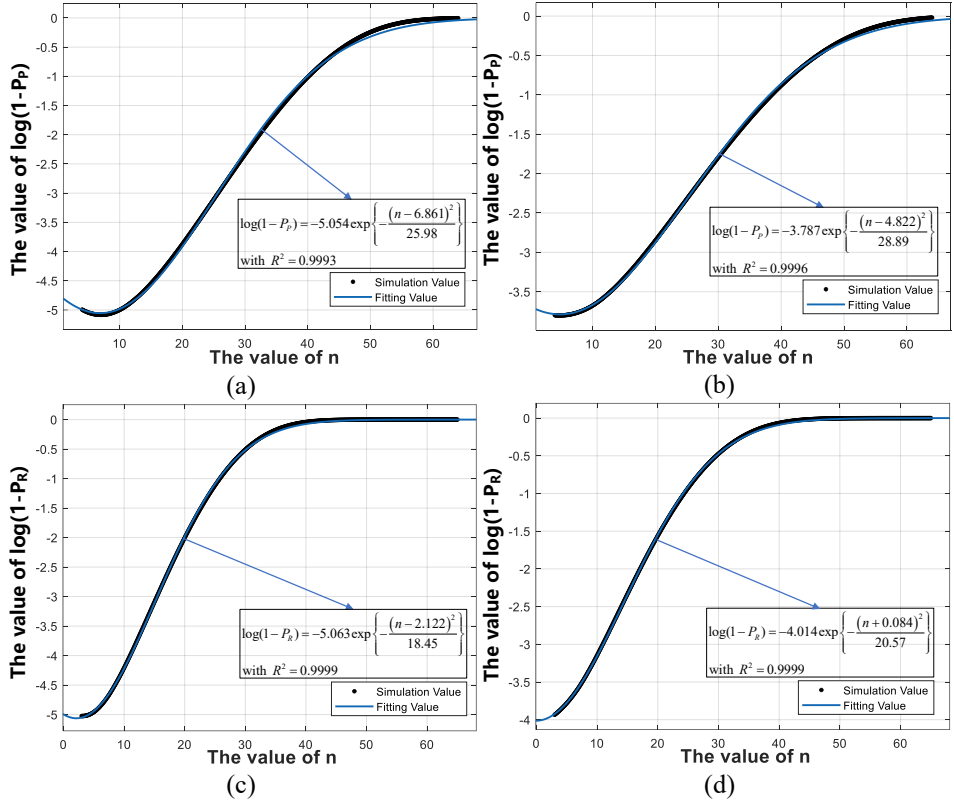

**Fig. 9.** (a) The value of log(1-$P_P$) for PBFT with THz signals; (b) The value of log(1-$P_P$) for PBFT with mmWave signals; (c) The value of log(1-$P_R$) for RAFT with THz signals; (d) The value of log(1-$P_R$) for RAFT with mmWave signals.

For **mmWave signals**, for both PBFT and RAFT, Fig.8 (b) shows that mmWave has a throughput property similar to THz, roughly in the order of $10^{15}$-$10^{16}$. This result indicates that the throughput of mmWave signals is inferior to that of THz signals, because mmWave signals result in higher consensus latency.

*C. Simulation of Reliability Gain*

Since the curve of consensus success rate about the number of nodes is difficult to be expressed by a concise mathematical formula, we explore the relationship between the logarithmic value of consensus success rate and the number of nodes, to enable researchers to better mine its law and promote the better deployment of wireless consensus networks in practical scenarios. This relationship is called *reliability gain*. It can reflect the ultimatum of wireless consensus networks and simplify the quantitative relationship between consensus success rate and the number of nodes.

In this paper, $z$=4 dB, and $\gamma$=5 nodes/m² are taken as examples, and the correlation between *reliability gain* and the number of nodes is shown in Fig.9. By fitting the curve of logarithmic values, we find a surprising phenomenon that the fitting curve perfectly accords with a Gaussian-like distribution. The *reliability gain* of wireless PBFT and RAFT networks are consistent with this rule for both THz and mmWave signals.

In Fig. 9 (a), (b), (c) and (d), we mark the parameters of the four fitting Gaussian-like distribution curves and the $R^2$ values (coefficient of determination, the closer the value is to 1, the better the fitting effect is). The $R^2$ values of these four fitting curves are all very close to 1, which further emphasizes the reliability of this law. However, how to determine the parameters of the fitting Gaussian-like distribution curve through mathematical derivation, as well as the inherent complex basic principle of the *reliability gain* satisfying the Gaussian-like distribution, requires further study and exploration.

*D. Simulation of Energy Consumption*

In this simulation part, we respectively study the influence of the number of nodes $n$ and consensus success rate on consensus *energy consumption*. (The following unit: 1aJ=$10^{-18}$J)

*1) The Number of Nodes vs Energy Consumption*

The number of nodes in wireless consensus networks represents the scalability of blockchain systems. In general, we want the system to be as scalable as possible. However, as the number of nodes increases, more communication processes will be added to the consensus process, leading to more energy consumption. Therefore, it is necessary and instructive to study the energy consumption of wireless PBFT and RAFT networks at various stages from the number of nodes.

Through the latency simulation, we find that the values of $z$ and $\gamma$ have little influence on the latency. And energy consumption is closely related to latency, so we take the average latency in this simulation.

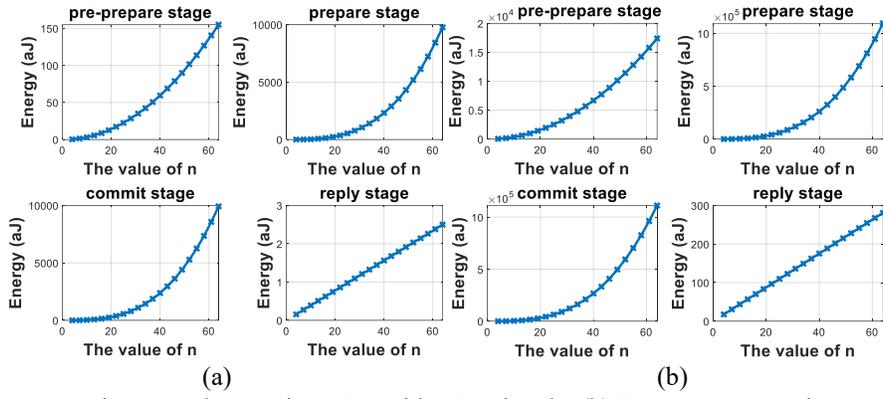

**Fig. 10.** (a) Energy consumption at each stage in PBFT with THz signals; (b) Energy consumption at each stage in PBFT with mmWave signals.

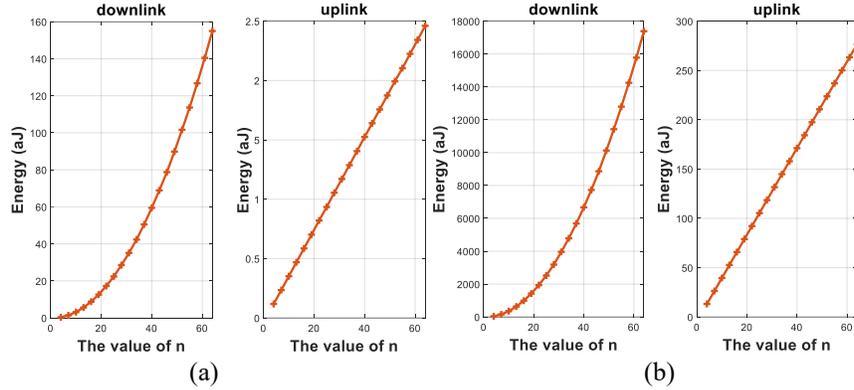

**Fig. 11.** (a) Energy consumption at each stage in RAFT with THz signals; (b) Energy consumption at each stage in RAFT with mmWave signals.

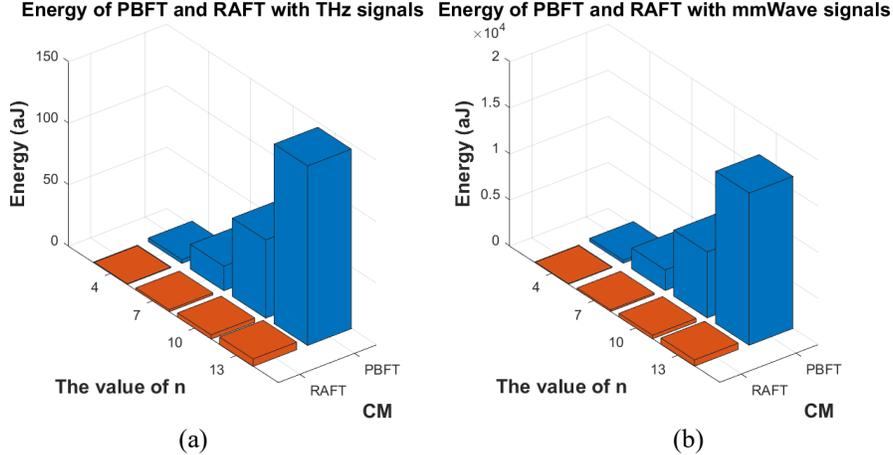

**Fig. 12.** (a) Energy consumption of PBFT and RAFT with THz signals; (b) Energy consumption of PBFT and RAFT with mmWave signals.

First, under THz and mmWave signals, we simulate the energy consumption of wireless PBFT networks at each stage, and results are shown in Fig. 10 (a) and (b). We find that in the first three stages (*pre-prepare*, *prepare*, *commit*) of the PBFT, energy consumption increases polynomially concerning the number of nodes, because of the complex communication between nodes. While in the fourth stage (*reply*), energy consumption increases linearly concerning the number of nodes. In addition, the energy consumption of mmWave signals is two orders of magnitude higher than that of THz signals, which is closely related to the latency of both.

Second, we simulate the energy consumption of wireless RAFT networks at the *downlink* and *uplink* stage in the environment of THz and mmWave signals, and results are shown in Fig. 11 (a) and (b). The energy consumption of *downlink* communications is similar to the first stage (*pre-prepare*) of PBFT, which is a quadratic upward trend about the number of nodes. Meanwhile, the difference in energy consumption between THz and mmWave signals in wireless RAFT networks is consistent with that in wireless PBFT networks.

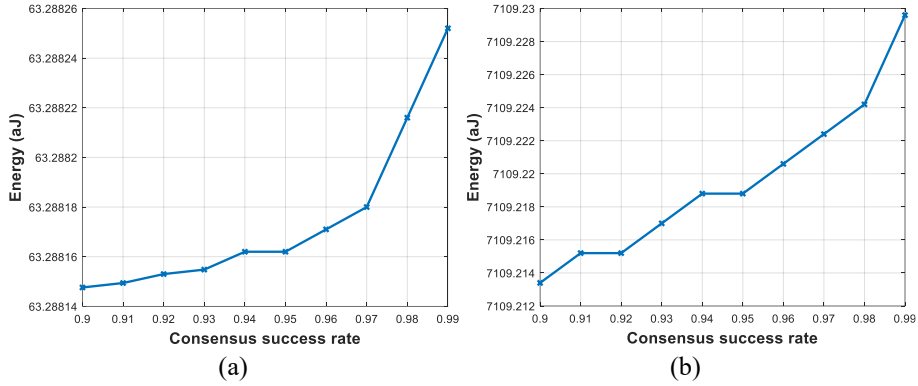

**Fig. 13.** (a) Energy consumption of PBFT with THz signals; (b) Energy consumption of PBFT with mmWave signals.

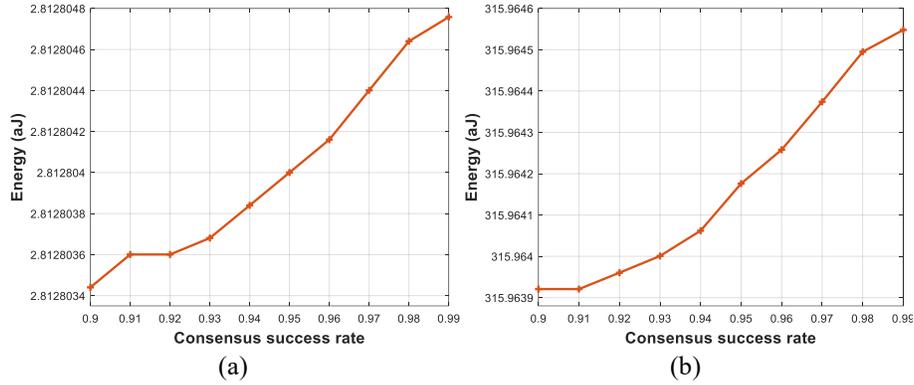

**Fig. 14.** (a) Energy consumption of RAFT with THz signals; (b) Energy consumption of RAFT with mmWave signals.

Third, we set the number of nodes as 4,7,10, and 13 to simulate the overall energy consumption of wireless PBFT and RAFT networks under THz and mmWave signals, and the results as shown in Fig.12 (a) and (b). These two diagrams clearly show the difference in energy consumption between the two CMs. The energy consumption of wireless PBFT networks is significantly higher than that of wireless RAFT networks.

*2) Consensus Success Rate vs Energy Consumption*

In addition to the influence of the number of nodes on energy consumption, the relationship between consensus success rate and the energy consumption is also worth exploring. This relationship illustrates an important question, namely, how much energy is needed to make wireless consensus networks work properly?

The success of CMs in reaching consistency is the result of all the stages working together, thus, this part shows the total energy consumption. We fix the number of nodes $n$=10, and explore the energy consumption with consensus success rate from 0.9 to 1, because only a high consensus success rate is meaningful for distributed systems.

The simulation results of the wireless PBFT network are shown in Fig. 12 (a) and (b). On the one hand, these results show that energy consumption increases slowly as the success rate of PBFT CM increases from 0.9 to 1. When the consensus success rate is closer to 1, the increasing trend of energy consumption is more obvious. On the other hand, the simulation also reveals that the energy consumption of the THz signal is two orders of magnitude lower than that of the mmWave signal, because the THz signal can provide more bandwidth resources and make the latency shorter.

Additionally, the simulation results of the wireless RAFT network can be viewed in Fig. 14 (a) and (b), which show that RAFT has a similar pattern to PBFT. Furthermore, the results also indicate that RAFT has an advantage over PBFT in terms of energy consumption. This is due to the simpler communication flow and shorter consensus latency of RAFT.

## VI. CONCLUSION

In this paper, the performance of non-ideal wireless PBFT and RAFT networks with mmWave and THz signals is investigated and analyzed. The study includes various performance metrics such as *consensus success rate*, *latency*, *throughput*, *reliability gain*, and *energy consumption*. The paper first reviews the PBFT and RAFT CMs and derives the transmission success rate of mmWave and THz signals in RF and FS models. Then, the theoretical calculation methods of the above performance metrics are derived, and numerical simulations are carried out to validate the results. Moreover, we derive the maximum distance between any two nodes in a network, called the *active distance* of CMs, which can make PBFT and RAFT consensus inevitably successful.

Through literature review, we find this is the first work to study the performance of blockchain consensus networks in mmWave and THz signals environment, which provides a valuable reference for the practical deployment of these technologies in 6G communications. In the future, it is expected to use the above performance analysis to design blockchain CMs with more practical features and functions such as ultra-reliable, low latency and low energy consumption.